\documentclass{svjour3}                     % onecolumn (standard format)

\usepackage{amssymb}
\usepackage{graphicx}
\usepackage{natbib}
\bibpunct{(}{)}{;}{a}{}{,}
\journalname{SSRv}

\begin{document}

\title{Clusters of galaxies: beyond the thermal view}

\author{
J.S.~Kaastra \and
A.M.~Bykov \and
S.~Schindler \and
J.A.M.~Bleeker \and
S.~Borgani \and
A.~Diaferio \and
K.~Dolag \and
F.~Durret \and
J.~Nevalainen \and
T.~Ohashi \and
F.B.S.~Paerels \and
V.~Petrosian \and
Y.~Rephaeli \and
P.~Richter \and
J.~Schaye \and
N.~Werner
}

\authorrunning{Kaastra, Bykov, Schindler et al.}

\institute{
J.S. Kaastra \at 
   SRON Netherlands Institute for Space Research, 
   Sorbonnelaan 2, 3584 CA Utrecht, the Netherlands \\
   Astronomical Institute, Utrecht University, P.O. Box 80000, 
   3508 TA Utrecht, The Netherlands \\
   \email{j.kaastra@sron.nl}
\and
A.M. Bykov \at 
   A.F. Ioffe Institute of Physics and Technology, St. Petersburg,
   194021, Russia \\
   \email{byk@astro.ioffe.ru}
\and
S. Schindler \at
   Institute for Astro- and Particle Physics,
   University of Innsbruck, Technikerstr. 25, 6020 Innsbruck, Austria \\
   \email{Sabine.Schindler@uibk.ac.at}
\and 
J.A.M. Bleeker \at 
   SRON Netherlands Institute for Space Research, 
   Sorbonnelaan 2, 3584 CA Utrecht, the Netherlands \\
   Astronomical Institute, Utrecht University, P.O. Box 80000, 
   3508 TA Utrecht, The Netherlands \\
   \email{j.a.m.bleeker@sron.nl}
\and
S. Borgani \at
   Department of Astronomy, University of Trieste, 
   via Tiepolo 11, I-34143 Trieste, Italy\\
   INAF -- National Institute for Astrophysics, Trieste, Italy \\  
   INFN -- National Institute for Nuclear Physics,
   Sezione di Trieste, Italy \\
   \email{borgani@oats.inaf.it}
\and
A. Diaferio \at
   Dipartimento di Fisica Generale ``Amedeo Avogadro'', 
   Universit\`a degli Studi di Torino,
   Via P. Giuria 1, I-10125, Torino, Italy \\
   Istituto Nazionale di Fisica Nucleare (INFN), 
   Sezione di Torino, Via P. Giuria 1, I-10125, Torino, Italy \\
   \email{diaferio@ph.unito.it} 
\and
K. Dolag \at 
   Max-Planck-Institut f\"ur Astrophysik,
   P.O. Box 1317, D-85741 Garching, Germany\\
   \email{kdolag@mpa-garching.mpg.de}
\and
F. Durret \at 
   Institut d'Astrophysique de Paris, CNRS, UMR~7095,
   Universit\'e Pierre et Marie Curie,
   98bis Bd Arago, F-75014 Paris, France \\
   \email{durret@iap.fr}
\and
J. Nevalainen \at
   Observatory, P.O. Box 14, 00014 University of Helsinki, 
   Helsinki, Finland \\
   \email{jnevalai@astro.helsinki.fi}
\and
T. Ohashi \at 
   Department of Physics, School of Science, Tokyo Metropolitan University, 
   1-1 Minami-Osawa, Hachioji, Tokyo 192-0397, Japan \\
   \email{ohashi@phys.metro-u.ac.jp}
\and
F.B.S. Paerels \at 
   Department of Astronomy and Columbia Astrophysics Laboratory,
   Columbia University, 550 West 120th Street, New York, NY 10027, USA \\
   SRON Netherlands Institute for Space Research,
   Sorbonnelaan 2, 3584 CA Utrecht, the Netherlands \\
   \email{frits@astro.columbia.edu}
\and
V. Petrosian \at 
   Department of Applied Physics, 
   Stanford University, Stanford, CA, 94305 \\
   Kavli Institute of Particle Astrophysics and Cosmology, 
   Stanford University, Stanford, CA, 94305 \\
   \email{vahep@stanford.edu}
\and
Y. Rephaeli \at
   School of Physics \& Astronomy, Tel Aviv University, 
   Tel Aviv, 69978, Israel\\
   Center for Astrophysics and Space Sciences, University of 
   California, San Diego, La Jolla, CA\,92093-0424 \\ 
   \email{yoelr@wise.tau.ac.il}
\and
P. Richter \at 
   Institut f\"ur Physik, Universit\"at Potsdam,
   Am Neuen Palais 10, D-14469 Potsdam, Germany\\
   \email{prichter@astro.physik.uni-potsdam.de}
\and
J. Schaye \at 
   Leiden Observatory, Leiden University, P.O. Box 9513, 
   2300 RA Leiden, The Netherlands \\
   \email{schaye@strw.leidenuniv.nl}
\and
N. Werner \at 
   SRON Netherlands Institute for Space Research, 
   Sorbonnelaan 2, 3584 CA Utrecht, the Netherlands \\
   \email{n.werner@sron.nl}
}

\titlerunning{Introduction}

\date{Received: 7 September 2007; Accepted: 20 December 2007}

\maketitle

\begin{abstract}

We present the work of an international team at the International Space Science
Institute (ISSI) in Bern that worked together to review the current
observational and theoretical status of the non-virialised X-ray emission
components in clusters of galaxies. The subject is important for the study of
large-scale hierarchical structure formation and to shed light on the "missing
baryon" problem. The topics of the team work include thermal emission and
absorption from the warm-hot intergalactic medium, non-thermal X-ray emission in
clusters of galaxies, physical processes and chemical enrichment of this medium
and clusters of galaxies, and the relationship between all these processes. One of the
main goals of the team is to write and discuss a series of review papers on this
subject. These reviews are intended as introductory text and reference for
scientists wishing to work actively in this field. The team consists of sixteen
experts in observations, theory and numerical simulations.

\end{abstract}

\section{Scientific rationale of the project}

Clusters of galaxies are the largest gravitationally bound structures in the
Universe. Their baryonic composition is dominated by hot gas that is in
quasi-hydrostatic equilibrium within the dark matter dominated gravitational
potential well of the cluster. The hot gas is visible through spatially extended
thermal X-ray emission, and it has been studied extensively both for assessing
its physical properties and also as a tracer of the large-scale structure of the
Universe.

Clusters of galaxies are not isolated entities in the Universe: they are
connected through a filamentary cosmic web. Theoretical predictions indicate the
way this web is evolving. In the early Universe most of the gas in the web was
relatively cool ($\sim 10^4$~K) and visible through numerous absorption lines,
designated as the so-called Ly$\alpha$ forest. In the present Universe, however,
about half of all the baryons are predicted to be in a warm phase
($10^5-10^7$~K), the Warm-Hot Intergalactic Medium (WHIM), with temperatures
intermediate between the hot clusters and the cool absorbing gas causing the
Ly$\alpha$ forest.

The X-ray spectra of clusters are dominated by the thermal emission from the hot
gas, but in some cases there appears to be evidence for hard X-ray tails or soft
X-ray excesses. Hard X-ray tails are difficult to detect, and one of the topics
for the team is a discussion on the significance of this detection (yet
contradictory) in existing and future space experiments. Various models have
been proposed to produce these hard X-ray tails, and our team reviews these
processes in the context of the observational constraints in clusters. 

While in some cases soft excesses in clusters can be explained as the low-energy
extension of the non-thermal hard X-ray components mentioned above, there is
evidence that a part may also be due to thermal emission from the WHIM.
The signal seen near clusters then originates in the densest and hottest parts
of the WHIM filaments, where the accelerating force of the clusters is highest
and heating is strongest. A strong component of this emission is line radiation
from highly ionised oxygen ions, and the role of this line emission and its
observational evidence will be reviewed. 

WHIM filaments not only can be observed because of their continuum or line
emission, but also through absorption lines if a sufficiently strong continuum
background source is present. The evidence for absorption
in both UV and X-ray high-resolution spectra is discussed. Future space missions will
be well adapted to study these absorption lines in more detail. 

In particular in absorption lines the lower density parts between clusters
become observable. In these low density regions of the WHIM not only collisional
ionisation but also photo-ionisation is an important process. In general, the
physics of the WHIM is challenging due to its complexity since there are many
uncertain factors including the heating and cooling processes, the chemical
enrichment, the role of supernova-driven bubbles or starburst winds,
ram-pressure stripping, the role of shocks, magnetic fields, etc. More detailed
(and sophisticated) hydrodynamical simulations with state-of-the-art spatial
(and temporal) resolution are required in order to follow the impact of some
(if not all) of these important processes. In particular chemical enrichment is
an important process to consider as it leads to many observable predictions. We
review the various physical processes relevant for the WHIM, the methods that
are used to simulate this and the basic results from those models.

\section{Timeliness of this work}

The first detections of non-virialised components in or between clusters of
galaxies such as thermal emission or absorption from the WHIM or the presence of
hard X-ray tails have now been made. There are several initiatives for new
space missions to study the physics of the WHIM, either in emission or
absorption, from the USA, Japan, Italy and The Netherlands, and new missions for
studying hard X-ray tails are being designed in France, Japan and the USA.
Significant theoretical progress is being made in this field, also thanks to the
enormous leap in computing power for numerical models. In addition to the
considerable intrinsic interest in the astrophysics of clusters, these systems
are fundamental probes of the underlying cosmology and of the large-scale
structure. This is the time to put the expertise of observers and theoreticians
in different fields together, in order to review our current knowledge and make
it available to the community in a self-contained and comprehensive -- yet
concise -- review volume.

\section{Organisation of the team work}

Our team consisted of sixteen members; two members unfortunately could not
attend both meetings; for one participant a replacement could be found. The
people involved in this international team were Xavier Barcons (Spain), Johan
Bleeker (Netherlands, co-organiser), Andrei Bykov (Russia, co-organiser),
Stefano Borgani (Italy), Antonaldo Diaferio (Italy), Klaus Dolag (Germany),
Florence Durret (France), Jelle Kaastra (Netherlands, organiser), Jukka
Nevalainen (Finland), Takaya Ohashi (Japan), Frits Paerels (USA), Vah\'e
Petrosian (USA), Yoel Rephaeli (Isra\"el), Philipp Richter (Germany), Joop
Schaye (Netherlands), Sabine Schindler (Austria, co-organiser), and Norbert
Werner (Netherlands). In addition a few people outside the team helped in
writing the review papers or even have taken up the lead in writing these
papers: Serena Bertone (UK), Chiara Ferrari (Austria), Federica Govoni (Italy),
Dunja Fabjan (Italy), Luca Tornatore (Italy), and Rob Wiersma (Netherlands).

Our team started with a one week meeting at the end of October 2006 at ISSI in
Bern. Each team member introduced a topic, and in the following discussion the
relevant issues for the review were collected. After the discussions, subteams
were formed around the different topics, and the outlines of the different
chapters as presented in the present volume were defined. Each team member
participated in several subteams and papers, either as first author or co-author.
In the time between the first and the second meeting, several subteams were able
to work out these drafts further. At the second meeting, the progress was
reviewed, comments were made on the drafts and the remaining time was used to
work further on the drafts. 

After the second meeting, the draft papers were finished by the authors and sent
to two internal referees from the team who are non-experts in the field, in
order to check the comprehensibility of the papers and of course for the normal
work of any referee. All papers were then reviewed by an anonymous, external and
expert referee. That there was a lively discussion and interaction is proven by the
more than 700 emails that were exchanged between the authors and the editor between
July and December 2007.

\section{About the title of the book}

Our series of review papers both appears as a special volume of Space Science Reviews,
as well as in the form of a book published by Springer. One of the hardest items to
resolve was to find a catchy title for the book. At our last meeting in June 2007 we
had a long and lively discussion about it without a clear outcome. Only shortly before 
the deadline we converged to the title of the book (which was taken also as the title of
this chapter / paper). While discussing what is beyond the thermal view, we spend of
course also quite some time on thermal emission. It is obvious that without a thorough 
understanding of the thermal aspects of the cluster gas, it is not well possible to go
beyond. But in this series of reviews we hope to demonstrate that cluster physics is
a rich subject, and that there is more beyond galaxies, dark matter, and ``just'' hot
gas with only one interesting parameter, its temperature (well, and its emission measure),
needed to estimate cluster masses. We go beyond this, and present non-thermal emission, 
shocks, magnetic fields, chemical enrichment, and all the other interesting processes 
shown on the front cover of the book. Also, we go beyond the -- somewhat diffuse -- 
physical boundaries of clusters to the warm-hot intergalactic medium with all its
interesting astrophysical aspects. We hope that our book and the individual papers
help the readers also to go beyond what is written here, and we invite them
to explore the topics of this book further and to advance our understanding of them. 

\section{Final remarks}

The team members look back to an interesting and rewarding project. The time
spent at ISSI has been successful, both from a point of view of the project, the
interactions and the atmosphere surrounding the project. The help and assistance
of ISSI and its staff members has been invaluable for this work.

\begin{figure}
\smallskip
\begin{center}
\includegraphics[angle=-0,width=\hsize]{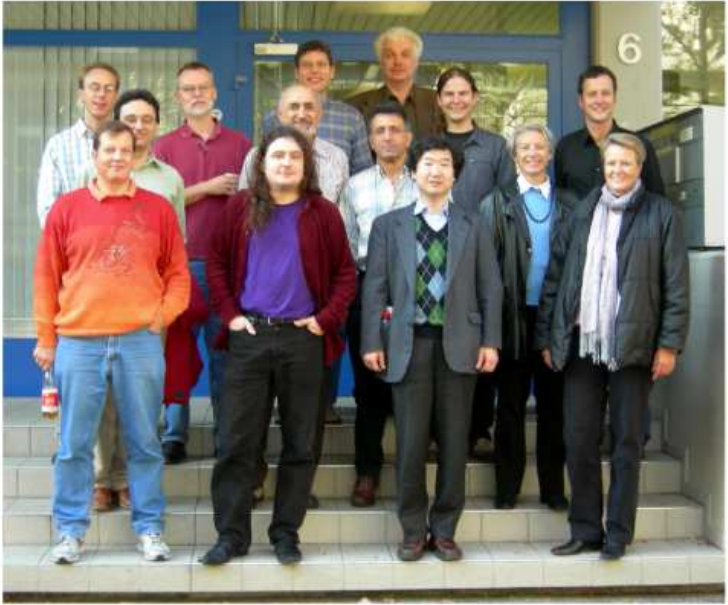}
\includegraphics[angle=-0,width=0.75\hsize]{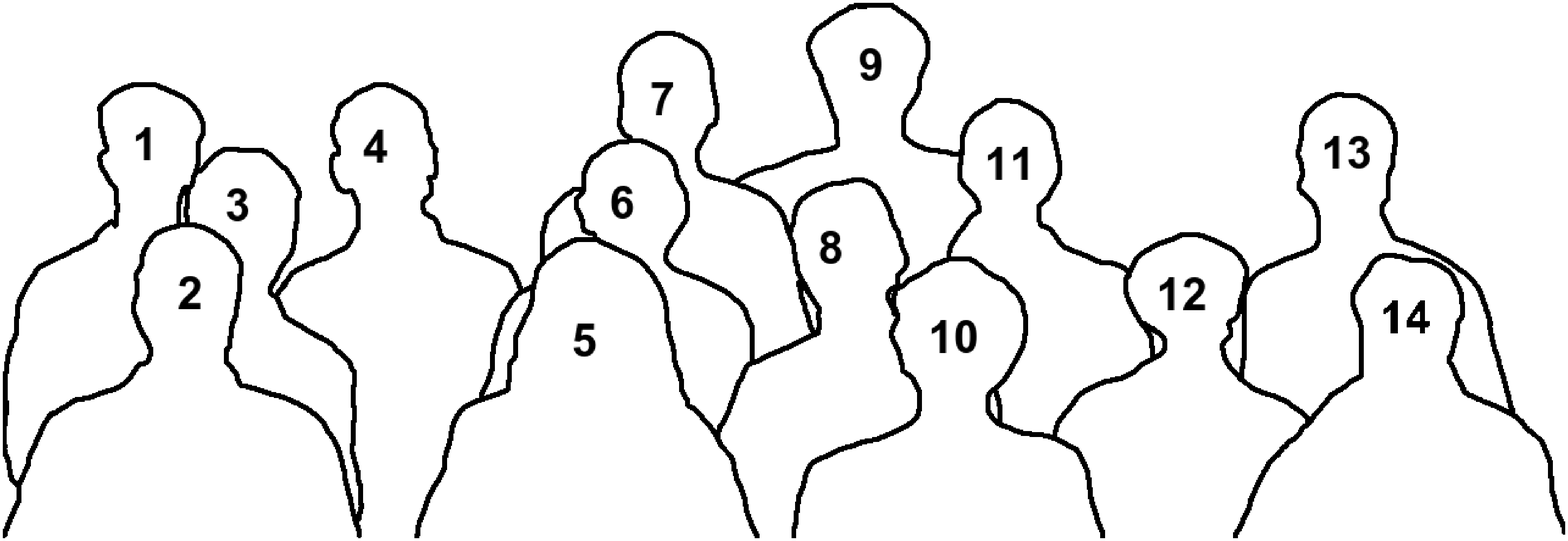}
\caption{The team members in front of the ISSI building at the first
meeting: Joop Schaye (1),
Klaus Dolag (2), Antonaldo Diaferio (3), Frits Paerels (4), Jukka Nevalainen
(5), Vah\'e Petrosian (6), Jelle Kaastra (7), Yoel Rephaeli (8),
Andrei Bykov (9), Takaya Ohashi (10), Norbert Werner (11), Florence Durret
(12), Philipp Richter (13) and Sabine Schindler (14).}
\label{fig:group}
\end{center}
\end{figure}

\begin{acknowledgements}
The authors thank ISSI (Bern) for support of the team ``Non-virialized X-ray
components in clusters of galaxies''. SRON is supported financially by NWO, the
Netherlands Organization for Scientific Research. 
A.M.B. acknowledges the RBRF grant 06-02-16844,
a support from RAS Presidium Programs, and a support from
NASA ATP (NNX07AG79G).
S.S. acknowledges financial support by the
Austrian Science Foundation (FWF) through grants P18523-N16 and
P19300-N16, by the Tiroler Wissenschaftsfonds and through the
UniInfrastrukturprogramm 2005/06 by the BMWF.
A.D. gratefully acknowledges
support from the PRIN2006 grant ``Costituenti fondamentali dell'Universo'' of
the Italian Ministry of University and Scientific Research and from the INFN
grant PD51.
F.D. acknowledges support from CNES. 
J.N. acknowledges support from the Academy of Finland. 
F.P. acknowledges
support from the Dutch Organization for Scientific Research NWO.
N.W. acknowledges support by the Marie Curie EARA Early Stage Training visiting
fellowship. 
\end{acknowledgements}

\end{document}